# Spectroscopic signature of phosphate crystallization in Erbium-doped optical fibre preforms

R. Peretti[1], A.M. Jurdyc[1], B. Jacquier[1], W. Blanc[2,*] and B. Dussardier[2]

[1] *Université de Lyon, Université Lyon 1, CNRS/LPCML, 69622 Villeurbanne, France*

[2] *Université de Nice-Sophia Antipolis, LPMC CNRS UMR6622, Parc Valrose, 06108 Nice Cedex 2, France*

*Corresponding author: wilfried.blanc@unice.fr

Abstract: In rare-earth-doped silica optical fibres, the homogeneous distribution of amplifying ions and part of their spectroscopic properties are usually improved by adding selected elements, such as phosphorus or aluminum, as structural modifier. In erbium ion ($Er^{3+}$) doped fibres, phosphorus preferentially coordinates to $Er^{3+}$ ions to form regular solvation shells around it. However, the crystalline structures described in literature never gave particular spectroscopic signature. In this article, we report emission and excitation spectra of $Er^{3+}$ in a transparent phosphorus-doped silica fibre preform. The observed line features observed at room and low temperature are attributed to $ErPO_4$ crystallites.

## 1. Introduction

Most optical fibres are made of silica, because of the many attractive properties of this glass: high mechanical strength, low thermal expansion, low refractive index, chemical durability, etc.

Rare-earth (RE) –doped optical fibres are now widely used in amplifiers for optical telecommunications, in lasers for mechanical processing as well as for metrology or medecine applications. Among the RE elements series, trivalent erbium ions ($Er^{3+}$) is especially interesting because its $^4I_{13/2}$-$^4I_{15/2}$ transition centered at 1.54 µm coincides with the lowest attenuation window of silica–based fibres. This has allowed for the development of high density, long-haul optical telecommunications networks based on Erbium Doped Fibre Amplifiers (EDFAs) [1,2].

Low solubility of RE ions in silica glass was a main concern at the early stage of the development of the active media. Indeed, in pure silica glass, RE ions can easily make clusters. In those glassy areas with a high RE ions concentration, the probability of the energy transfer mechanism between RE ions increases dramatically leading to the concentration quenching mechanism [3]. This degrades the performance of devices based on RE-doped fibres, such as EDFAs. To overcome this drawback, it was proposed to dope silica with network modifying cations, such as $Al^{3+}$ or $P^{5+}$, to dissolve RE ions in the silica modified network. Nowadays EDFAs are based on Al-codoped fibres because they have broader and relatively smoother gain spectrum compared to P-doped fibres. However, P-codoped fibres are interesting because they induce larger emission cross sections [4], and efficient non-radiative energy transfers in Yb-Er codoped power amplifiers [5]. The spectroscopic properties of neodymium ions ($Nd^{3+}$) were also improved by P-codoping [6], even at low $P_2O_5$ content (< 1mol%) [7]. It was proposed that P co-dopants preferentially coordinate to $Nd^{3+}$ ions to form a solvation shell structure. Such conclusions were reported only twice for erbium ions in optical fibre preforms prepared by the Modified Chemical Vapor Deposition (MCVD) process [8,9]. Reports tend to demonstrate that $Er^{3+}$ ions are inserted in a locally well ordered phase such as $ErPO_4$ [8,9]. This ordered environment would avoid any RE ions clusterings. However, no signature of such crystalline structures in spectroscopic characteristics was ever reported in those sample with high $P_2O_5/Er_2O_3$ ratio (10 and 500 in Refs 9 and 8, respectively). On the contrary, their $Er^{3+}$ emission spectra were always composed of broad emission bands, as expected from an amorphous matrix.

In this paper, we investigate a P- and Er-codoped germano-silicate fibre preform made by the MCVD. Compared to previously reported results, we investigated a relatively low $P_2O_5/Er_2O_3$ ratio equal to 2. Moreover, $P_2O_5$ and $Er_2O_3$ concentrations were chosen according to fibre fabrication considerations and laser applications, respectively. We present for the first time emission and excitation spectra from $Er^{3+}$ ions showing narrow lines in a glass, both at room and low temperature, characteristic of the local ordering around the $Er^{3+}$ ions. These features are related to those observed in $ErPO_4$ materials. The reported results provide a better understanding of the close environment of RE ions in glass. Such optical fibres with a crystallized-like $Er^{3+}$ properties is interesting for developing of new lasers and would be of importance for the quantum memories [10].

## 2. Experimental

The silica preform was prepared by the usual MCVD process [11]. Germanium (for index rising of the core material) and phosphorus were added in the core layers during preform fabrication. The final $GeO_2$ and $P_2O_5$ concentrations in the collapsed preform, measured through EPMA analyses, were estimated to be about 3 and 0.1 mol%, respectively. No aluminum was added. $Er^{3+}$ ions were introduced through the well-known solution doping technique, using an alcoholic solution of $ErCl_3:6H_2O$ salt [12]. At the stage of core synthesis by MCVD, a porous silica layer is prepared, and further soaked with the solution. After drying of the solvent, the core layer is sintered down to a dense glass layer. Then the tube is collapsed into a solid rod, referred to as perform, at an elevated temperature higher than 1800 °C. Fibers were drawn to 125 μm by stretching the preform in a fiber-drawing tower at temperatures higher than 2000 °C under otherwise normal conditions. The attenuation at 978 nm in the drawn optical fibre was measured to be 30 dB/m. From this value the $Er_2O_3$ concentration is estimated to be 600 ppm mol. The transverse $P_2O_5/Er_2O_3$ ratio through the center of the preform is assumed to be constant [13]. The refractive index difference between the core and the cladding is $4.10^{-3}$. The core diameter in the preform is 1 mm. The length of the sample is 3 mm and the diameter of the preform is 1 cm. The sample was polished with good optical quality before experiments.

Emission and excitation spectra were recorded at room and very low (~1.5 K) temperatures. A laser diode from JDSU emitting at 980 nm with an output power of around 100 mW was used as excitation source to record emission spectra at room temperature. Experiments at 1.5 K and excitation experiments were recorded by using a narrow tunable laser diode Tunics in the 1480-1580 nm wavelength range. In glasses, high excitation energy (980 nm and 1519 nm) in any level gives the same emission spectra because all the sites are excited. Moreover, at room temperature there is no site selection for $Er^{3+}$ ions in silicate glasses. The laser power and beam waist reaching the sample was ~1 mW and ~100 μm, respectively, so that all non-linear effects were negligible. We have checked that our results are position independent. The sample was placed inside a super fluid helium bath cryostat which ensured to reach temperature as low as 1.5 K. The luminescence was collected and focused on the entrance of a 1 m-focal length Jobin-Yvon U1000 infrared monochromator. The filtered luminescence was then detected by a high-sensitivity germanium-cooled detector from North Coast. We used a mechanical shutter from Uniblitz to modulate the luminescence by a square signal at around 15 Hz, and a lock-in amplifier RS830 from Stanford Research to improve the signal to noise ratio.

## 3. Results

The emission and excitation spectra at low temperature (1.5 K) are shown in Fig. 1 and 2, respectively. The $Er^{3+}$ $^4I_{13/2} \rightarrow {}^4I_{15/2}$ emission spectrum was obtained under a 1519 nm (6583 cm$^{-1}$) excitation. The excitation spectrum corresponds to the 1560 nm (6410 cm$^{-1}$) emission wavelength ($^4I_{13/2} \rightarrow {}^4I_{15/2}$ transition). Emission and excitation wavelengths were chosen to avoid any site selection. All the excitation spectra recorded at 1.5 K were insensitive to the selected emission wavelength, apart for the absolute emitted spectral optical density.

The emission and excitation spectra present narrow bands which are unusual for silicate glasses and reveal a strong difference compared to the usual broad band spectra of $Er^{3+}$ ions in glassy environment [14]. We assume that at 1.5 K only the lowest levels of both the $^4I_{13/2}$ and $^4I_{15/2}$ manifolds are populated, split by the local field of the ions surrounding the $Er^{3+}$ ions. Therefore, the emission peaks should reflect the transitions from the lowest state of the $^4I_{13/2}$ to the $^4I_{15/2}$ manifolds. Seven main emission peaks are identified at low temperature : 6529, 6501, 6480, 6393, 6328, 6303 and 6260 $cm^{-1}$, respectively. The uncertainty on the wavenumber position is ± 2 $cm^{-1}$. The positions of the energy levels of the $^4I_{15/2}$ manifold are deduced from the energy difference between the transitions of the emission peaks. They are reported in Table 1. The peaks in the excitation spectrum come from the transitions from the lowest ground state ($^4I_{15/2}$) to the states of the $^4I_{13/2}$ manifold. Ten main excitation peaks are identified and are reported in Table 1 : 6530, 6540, 6562, 6582, 6592, 6600, 6608, 6619, 6638 and 6668 $cm^{-1}$, respectively.

The emission and excitation spectra measured at room temperature are shown in Fig. 3 and 4, respectively. These spectra present also broad and narrow bands. Upon increasing of the temperature from 1.5 K to room temperature, peaks are observed at 6667, 6635, 6618, 6533, 6501, 6397, 6339, 6303 and 6254 $cm^{-1}$, respectively (Fig. 3). In particular, new peaks appeared at higher energy compared to low temperature emission peaks (Fig. 1). This effect can be related to the thermal population of the higher levels of the $^4I_{13/2}$ manifold. The excitation spectra measured at low and room temperature, respectively (Fig. 2 and 4), present some common features, taking into account the homogeneous linewidth broadening induced by the temperature. Room temperature emission and excitation spectra present a larger bandwidth compared to low temperature measurements, which is expected in the case of $Er^{3+}$ in glassy environment. Also at room temperature, the excitation is less selective than at low temperature, i.e. more sites are excited at the same time, increasing both intensity and bandwidth. Then, emission spectra change with temperature as $Er^{3+}$ ions in the glassy phase are excited with a higher probability than at low temperature, whereas at low temperature $Er^{3+}$ ions in crystalline phases are preferentially excited.

*Fig. 1*

*Fig. 2*

*Fig. 3*

*Fig. 4*

**4. Discussion**

Emission and excitation spectra, both at room and low temperature, are characterized by the presence of narrow lines. Such features are usually related to the crystallized environment of the rare-earth ions. This is the first time, to our knowledge, that such features are observed in silica preforms for optical fibres. In crystals, with low symmetry, the number of Stark levels in the $^4I_{15/2}$ and $^4I_{13/2}$ manifolds has been predicted theoretically to be 8 and 7 [15-17], respectively. According to the emission and excitation spectra recorded at low temperature, we identified 7 and 10 sub-levels for the $^4I_{15/2}$ and $^4I_{13/2}$ manifolds, respectively (Table 1). The greater maximum number of Stark levels for the $^4I_{13/2}$ manifolds tends to indicate that at least two crystal phases could be present in the sample.

*Table 1*

In previous work, EXAFS measurements were done on samples with composition similar to that of the present sample [9]. $Er^{3+}$ ions were found to be ordered with an $ErPO_4$ environment. Then, to correlate the excitation and emission peaks with a crystallized environment of $Er^{3+}$ ions in this sample, we compare our results with those obtained in four phosphates : phosphate glass (Kigre QE7) [18], $ErPO_4$ [15], $LuPO_4$:Er [16] and $YPO_4$:Er [17]. In the two last materials, $Er^{3+}$ ions substitute to $Lu^{3+}$ or $Y^{3+}$ ions, respectively. In all the selected materials, $Er^{3+}$ ions are located in phosphate environment.

The Stark splitting between the lowest and the highest sub-level of both the $^4I_{15/2}$ and $^4I_{13/2}$ manifolds are reported in Table 2. The Stark splitting values measured for both manifolds are close to those reported for $ErPO_4$, $LuPO_4$:Er and $YPO_4$:Er. On the contrary, the phosphate glass based on

$P_2O_5$, ZnO, $Al_2O_3$ and $Er_2O_3$ [19] has a much higher Stark splitting than all the other materials. The difference can be explained by the $C_{2v}$ site symmetry occupied by $Er^{3+}$ ions in the glass from Kigre [15] while it is $D_{2d}$ in the other materials [16-18]. This comparison reinforces the fact that $Er^{3+}$ ion in the P-doped preform lies in an environment close to $ErPO_4$.

*Table 2*

Predicted and observed energy positions of the Stark levels are reported in the Table 1. Among the visible peaks in the emission and excitation spectra measured at low temperature, six and seven energy positions can be identified to those of $XPO_4$ (*X*= Er, Lu or Y) for the $^4I_{15/2}$ and $^4I_{13/2}$ manifolds, respectively. One could note that two small peaks are also present in the emission spectrum around 6420 cm$^{-1}$ (Fig. 1). According to values reported in Table 1, these peaks are tentatively assigned to transitions from the fundamental Stark level of the $^4I_{13/2}$ manifold down to a Stark level in the $^4I_{15/2}$ ground manifold lying around ~100 cm$^{-1}$.

Unassigned lines in Table 1 correspond to energy positions lower than those attributed in the previous paragraph. As only pure silica leads to a lowest Stark splitting for both the $^4I_{15/2}$ and $^4I_{13/2}$ manifolds compare to P-doped silica [20], the unassigned lines could be attributed to $Er^{3+}$ ions located in silica-based crystal, i.e. not coordinated with phosphorus.

## 5. Conclusion

In silica glass, phosphorus ions are known to form solvation shell structures around rare-earth ions. However, this prediction was never observed spectroscopically. In this article, we demonstrate the spectroscopic signature of some crystalline environment of $Er^{3+}$ in a transparent phosphor germano-silicate preform. We observed at least two crystal phases and a glassy phase. By comparing our results with those of literature, we have evidenced that one of the crystal phases could be related to $ErPO_4$. This result is in accordance with previous observations and EXAFS results on similar samples.

This kind of line structured spectrum for rare earth in transparent materials could be useful to built laser fibre, taking benefit both of the advantages of silica as fibre material and the narrow line of the rare-earth ions in the crystallized structure. Such a narrow linewidth is also interesting for the quantum memories. Finally, this study provides more insight on the fine structure of the close environment of RE ions in silica-based optical fibres.

**Acknowledgements**

The authors thank Michèle Ude and Stanislaw Trzésien for samples preparation. LPMC is a member of the GIS 'GRIFON' (http://www.unice.fr/GIS/).

**Table captions**

Table 1 : Comparison of Stark splitting between our sample and theoretical (Th.) and experimental (Exp.) data from erbium-doped phosphates. Data in grey framework show level energy values obtained in this work that are equal or close to reported values.

Table 2 : Stark splitting between the lowest and the highest level of the $^4I_{15/2}$ and $^4I_{13/2}$ manifolds of $Er^{3+}$ ions in different materials at low temperature.

Table 1

| Manifold | This work (measurements at 1.5 K) | | ErPO$_4$ [15] | LuPO$_4$:Er [16] | | YPO$_4$:Er [17] | |
|---|---|---|---|---|---|---|---|
| | Emission | Excitation | Th. | Th. | Exp. | Th. | Exp. |
| $^4I_{15/2}$ | 0 | | 0 | -1 | 0 | 3 | 0 |
| | 28 | | 33 | 35 | 36 | 28 | 32 |
| | 49 | | 53 | 49 | 53 | 47 | 52 |
| | | | 105 | 100 | 98 | 121 | 115 |
| | 136 | | 145 | 132 | | 149 | 147 |
| | | | | | | 186 | 185 |
| | 201 | | | | | 204 | 206 |
| | 226 | | 230 | 229 | | | |
| | | | 247 | 246 | | 241 | 241 |
| | 269 | | | | | | |
| | | | 287 | 286 | | | |
| $^4I_{13/2}$ | | 6530 | | 6548 | 6535 | 6540 | |
| | | 6540 | | 6556 | 6544 | 6543 | |
| | | 6562 | | | | | |
| | | 6582 | | | | | |
| | | 6592 | | | | 6591 | |
| | | 6600 | | 6608 | 6602 | | |
| | | 6608 | | | | | |
| | | 6619 | | 6619 | 6615 | 6620 | |
| | | 6638 | | 6647 | 6641 | | |
| | | | | | | 6655 | |
| | | | | | | 6658 | |
| | | 6668 | | | | 6669 | |
| | | | | 6687 | 6682 | | |
| | | | | 6697 | 6695 | | |

Table 2

| Manifold | This work | Phosphate [18] | ErPO$_4$ [15] | LuPO$_4$ [16] | YPO$_4$ [17] |
|---|---|---|---|---|---|
| $^4I_{13/2}$ | 270 | 361 | 287 | 286 | 241 |
| $^4I_{15/2}$ | 138 | 257 | - | 149 | 129 |

**Figure captions**

Figure 1: Er$^{3+}$ $^4I_{13/2}$ → $^4I_{15/2}$ emission spectrum, measured at 1.5K under a 1519 nm (6583 cm$^{-1}$) excitation. Identified peaks (see texte and Table 1) are indicated with a star symbol '*'.

Figure 2: Excitation spectrum of the 1560 nm (6410 cm$^{-1}$) emission wavelength, measured at 1.5 K. Identified peaks (see texte and Table 1) are indicated with a star symbol '*'.

Figure 3: Er$^{3+}$ $^4I_{13/2}$ → $^4I_{15/2}$ emission spectrum, measured at room temperature under a 980 nm (1020 cm$^{-1}$) excitation. Identified peaks (see texte and Table 1) are indicated with a star symbol '*'.

Figure 4 : Excitation spectrum of the 1560 nm (6410 cm$^{-1}$) emission wavelength, measured at room temperature.

Figure 1

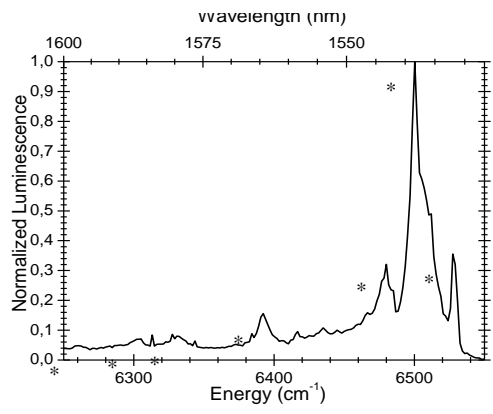

Figure 2

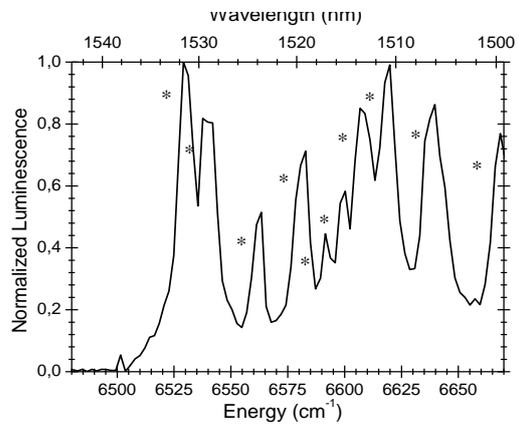

Figure 3

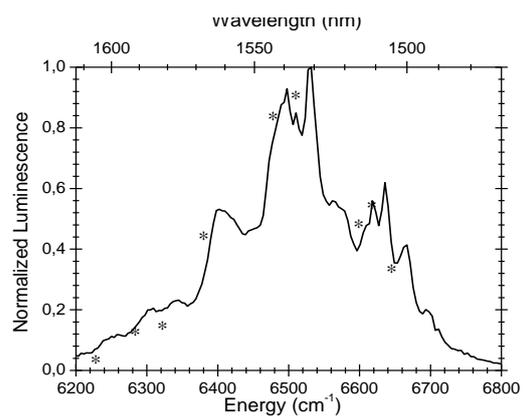

Figure 4

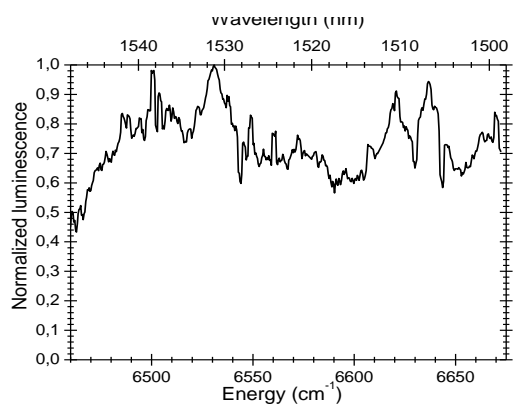